# Measurements of anisotropic g-factors for electrons in InSb nanowire quantum dots

Jingwei Mu[1,2], Shaoyun Huang[1], Ji-Yin Wang[1], Guang-Yao Huang[1], Xuming Wang[1], and H. Q. Xu[1,2,3,*]

[1] Beijing Key Laboratory of Quantum Devices, Key Laboratory for the Physics and Chemistry of Nanodevices, and Department of Electronics, Peking University, Beijing 100871, China

[2] Academy for Advanced Interdisciplinary Studies, Peking University, Beijing 100871, China

[3] Beijing Academy of Quantum Information Sciences, Beijing 100193, China

[*]To whom correspondence should be addressed. Email: hqxu@pku.edu.cn

(September 25, 2020)

**Abstract**

We have measured the Zeeman splitting of quantum levels in few-electron quantum dots (QDs) formed in narrow bandgap InSb nanowires via the Schottky barriers at the contacts under application of different spatially orientated magnetic fields. The effective g-factor tensor extracted from the measurements is strongly anisotropic and level-dependent, which can be attributed to the presence of strong spin-orbit interaction (SOI) and asymmetric quantum confinement potentials in the QDs. We have demonstrated a successful determination of the principal values and the principal axis orientations of the g-factor tensors in an InSb nanowire QD by the measurements under rotations of a magnetic field in the three orthogonal planes. We also examine the magnetic-field evolution of the excitation spectra in an InSb nanowire QD and extract a SOI strength of $\Delta_{so} \sim 180$ μeV from an avoided level crossing between a ground state and its neighboring first excited state in the QD.

## 1. Introduction

Semiconductor InSb nanowires are among the emerging, key materials for developments of solid-state based quantum device and quantum information technology. These nanomaterials have been employed in building semiconductor quantum dot (QD) spin qubits with fast operations of spin states by electrical means [1-3] and in constructing semiconductor-superconductor hybrid structures with potential applications towards topological quantum computing [4,5]. Strong spin-orbit interaction (SOI) and large g-factor have been exploited as the key physics properties of the nanomaterials in these applications [2-5]. Bulk InSb possesses an electron g-factor of −51 [6]. In low-dimensional InSb systems, such as InSb nanowire QDs, the magnitude of the electron g-factor is found to remain large but vary from level to level [7]. Due to a complex confinement potential profile and coupling of the spin degree of freedom of an electron to its orbital motion, the g-factor of electrons in a low-dimensional InSb quantum structure is expected to be strongly anisotropic. There have been studies showing that the g-factors in nanostructures, such as metal copper nanoparticles [8,9], semiconductor InAs QDs [10-17], Si QDs [18-20], InP nanowire QDs [21], Ge-Si core/shell nanowire QDs [22], InAs/InAlGaAs self-assembled QDs [23], and p-type GaAs/AlGaAs QDs [24], are anisotropic and/or level-dependent. Especially, with regards to semiconductor nanowire quantum structures, large orbital contributions to the electron g-factor have recently been theoretically identified [25] and experimentally demonstrated [17]. Some works on InSb nanowire QDs have also been reported [7,26,27,28]. However, full, solid experimental measurements of the anisotropic properties of the g-factor in InSb nanowire QDs have still not yet been carried out, although it is highly anticipated that the information obtained from such measurements is crucial for controlling of spin-orbit qubits [29,30] and topological Majorana zero modes [31] created in InSb nanowire quantum structures.

In this article, we report on low-temperature transport measurements of the Zeeman splitting of spin-1/2 electrons in single InSb nanowire QDs. The InSb nanowires are grown via metal organic vapor phase epitaxy and each QD is defined between the two metal contacts in an InSb nanowire via the naturally formed Schottky barriers. We extract the g-factors from the measured Zeeman splitting of spin-1/2 electrons at quantum levels of the QDs in the few-electron regime. We show the measurements performed for InSb nanowire QDs under the magnetic fields applied along the three axes of a coordinate system, in which one axis is set along the nanowire axis, and demonstrate that the g-factor of a spin-1/2 electron in these QDs is strongly anisotropic and level-dependent. The measurements are also performed for an InSb nanowire QD under 360°-rotations of an applied magnetic field in three orthogonal planes. These measurements allow us to determine the principal values and principal axes of the g-factor tensors of spin-1/2 electrons at quantum levels of the QD and thus provide a complete experimental demonstration for anisotropy and level dependence of the g-factor in the InSb nanowire QDs. To extract the strength magnitude of SOI in the InSb nanowire QDs, we also perform the measurements for the excitation spectra of an InSb nanowire QD as a function of the magnetic field applied perpendicular to the nanowire. From the anti-crossing of two neighboring quantum levels occupied by electrons with opposite spins in the excitation spectra, a SOI strength of $\Delta_{SO} \sim 180$ μeV is extracted.

## 2. Methods

The InSb nanowires employed in this work for device fabrication are grown on top of InAs nanowires by metal-organic vapor-phase epitaxy on an InAs (111)B substrate [32]. Previous studies showed that these InSb nanowires are in zincblende phase, grown along the [111] crystallographic direction and free from stacking faults [32]. The nanowires are transferred by a dry method onto a heavily n-doped Si substrate, capped with a 200-nm-thick layer of $SiO_2$ on top, with predefined Ti/Au bonding pads and markers. The Si substrate and the capped $SiO_2$ layer serve as a global back gate and a gate dielectric to the nanowires, respectively. Electron-beam lithography is used to define the source and drain contact areas on the nanowires. Then, in order to remove surface oxides and obtain clean metal-semiconductor interfaces, we etch the contact areas in a low concentration (~2%) ammonium polysulfide [$(NH_4)_2S_x$] solution at 40 °C for 1 min [33]. After the etching process, contact electrodes are fabricated by deposition of 5-nm-thick titanium and 90-nm-thick gold via electron-beam evaporation and lift off. In each fabricated device, a QD is defined in the InSb nanowire segment between the two contacts by naturally formed Schottky barriers [7,26]. Figure 1(a) shows a false-colored scanning-electron microscope (SEM) image of a representative fabricated device and figure 1(b) is a schematic cross-sectional view of the device structure and the measurement circuit setup.

The fabricated QD devices are studied by transport measurements in a dilution refrigerator equipped with a vector magnet. In this work, we will report the results of measurements for two representative InSb nanowire QD devices, which will be labeled as devices A and B. Device A is made from an InSb nanowire of ~85 nm in diameter with a contact spacing of ~150 nm and device B is made from an InSb nanowire of ~75 nm in diameter with the same (~150 nm) contact spacing. In the measurements, the back gate [cf. figure 1(b)] is used to control the chemical potential in the QDs and the magnetic field is applied in a direction with respect to the coordinate system with the X axis pointing along the nanowire axis as shown in figure 1(b).

## 3. Results and discussion

*3.1 Transport characteristics of the QD in device A*

Figure 1(c) shows the differential conductance $dI_{ds}/dV_{ds}$ measured for device A as a function of source-drain bias voltage $V_{ds}$ and back-gate voltage $V_{bg}$ (charge stability diagram) at zero magnetic field. Typical quantum transport characteristics of a single QD in the few-electron regime are observed in the measurements. The regular diamond-shaped low-conductance regions are the regions in which the number of electrons in the QD is well defined and the electron transport through the QD is blockade by Coulomb repulsion (Coulomb blockade effect). Since no more Coulomb diamond structure is observable at back-gate voltages $V_{bg}$ lower than −2.55 V, the QD is empty of electrons at $V_{bg}$ below −2.55 V. As a consequence, the electron numbers in the QD in the first three Coulomb blockade regions can be assigned [as indicated by integer numbers *n* in figure 1(c)]. The alternative changes in Coulomb diamond size seen in figure 1(c) are the traces of spin degeneracy and level quantization in the QD [34,35]. In the first Coulomb blockade region, the QD is occupied by one electron. For a second electron of the opposite spin to enter and then pass through the QD, an excess energy is required to overcome single electron charging energy $E_C$. Thus, from the vertical span of the first Coulomb diamond, a charging energy of $E_C$~5.4 meV is extracted. The total capacitance of the QD in this back-gate voltage region is $C_\Sigma$~29.7 aF. We can also extract, from the horizontal span of the first Coulomb diamond, a back-gate capacitance of $C_g$~2.2 aF, which together with the total capacitance $C_\Sigma$ gives a back-gate lever arm factor of α (= $C_g/C_\Sigma$)~0.075. In the second Coulomb blockade region, where the QD is occupied by two electrons of opposite spins, adding one more electron to the QD needs to overcome an electron addition energy which includes both single electron charging energy $E_C$ and level quantization energy $\Delta\varepsilon$ [7]. Thus, by assuming that $E_C$ remains unchanged, we can extract, from the vertical span of the second Coulomb diamond (a measure for the addition energy of the third electron to the QD), a level quantization energy of $\Delta\varepsilon\sim3.9$ meV. The quantization energy could also be estimated out from excited state conductance lines located outside of the Coulomb diamonds in the charge stability diagram. For example, along the line cut A in figure 1(c), the low-energy high conductance line involves a transition of quantum states from the spin-↑ doublet ground state $|D_\uparrow\rangle$ to the two-electron singlet ground state $|S\rangle$ and the high-energy high conductance line involves transitions of quantum states from the spin-↑ doublet ground state $|D_\uparrow\rangle$ to two-electron excited states $|S^*\rangle$ and $|T^*\rangle$ involving both the first and the second quantum level [36]. Thus, the energy difference between the two-electron singlet ground state and excited states can be extracted out from the half of the vertical distance in bias voltage between the two high-conductance lines, which yields $\Delta\varepsilon\sim3.8$ meV. This value is in good agreement with the estimation made above from the difference in addition energy between the first and second Coulomb diamonds.

*3.2. g-factor measurements for the first two quantum levels of the QD in device A*

Figure 2 shows the magnetic field evolutions of the first and second quantum levels of the QD in device A. Figure 2(a) displays the current $I_{ds}$ measured for device A at $V_{ds} = 0.2$ mV as a function of $V_{bg}$ and magnetic field $B_X$ applied along the X direction. Here, the spin states of subsequently filled electrons in the level are marked by arrows. As $B_X$ is increased from zero, the current peaks corresponding to the spin-↑ and spin-↓ states move apart. After converting the peak positions in energy using the back-gate lever arm factor determined above, an effective g-factor $g_{IX}^*$ of the first quantum level (level *I*) can be extracted from the energy difference $\Delta E(B_X)$ between the two spin states according to $\Delta E(B_X) = E_C + |g_{IX}^*\mu_B B_X|$, where $\mu_B$ is the Bohr magneton. Figure 2(b) shows the measured energy separations of the two spin states as a function of $B_X$. By a line fit to the measurements [red solid line in figure 2(b)], we extract a value of $g_{IX}^*\sim58.5$ for the g-factor of quantum level *I* in the QD. Similar measurements are performed with magnetic fields $B_Y$ and $B_Z$ applied along the Y and Z axes, and the corresponding g-factors, $g_{IY}^*\sim51.2$ and $g_{IZ}^*\sim55.5$, are extracted. The inset of figure 2(b) summarizes the three extracted g-factor values of the quantum level at magnetic fields applied along the X, Y and Z axes. Clearly, the three g-factor values are different, implying that the g-factor of the quantum level in the QD is anisotropic.

Figure 2(c) shows the differential conductance $dI_{ds}/dV_{ds}$ measured along the line cut A in figure 1(c) as a function of source-drain bias voltage $V_{ds}$ and magnetic field $B_X$. Here, the three high conductance stripes are observed. The upper one involves a transition of quantum states from the spin-↑ doublet ground state $|D_\uparrow\rangle$ to the two-electron singlet ground state $|S\rangle$ at a finite magnetic field and can be associated with the tunneling process of a spin-↓ electron through the QD. As a result, this high conductance stripe moves to higher energy with increasing $B_X$. Note that negative values of applied bias voltage $V_{ds}$ are given in figure 2(c). Since both the doublet and singlet ground states, $|D_\uparrow\rangle$ and $|S\rangle$, involve only the lowest quantum level in the QD, we can extract the electron g-factor of the quantum level in the QD from the slope of the high conductance stripe. The result is $g_{IX}^* \sim 57.3$, which is in good agreement with the corresponding g-factor extracted from the measurements shown in figures 2(a) and 2(b). Similar measurements are performed for the magnetic field applied along the Y and Z directions and the corresponding g-factors are extracted to be $g_{IY}^* \sim 50.8$ and $g_{IZ}^* \sim 55.6$, which are all in good agreement with the results extracted from the measurements as described in figures 2(a) and 2(b).

The lower two split high conductance stripes in figure 2(c) involve the QD state transitions from the spin-↑ doublet ground state $|D_\uparrow\rangle$ to three two-electron excited states $|S^*\rangle$, $|T_0^*\rangle$, and $|T_+^*\rangle$ involving both the first and the second quantum level in the QD [36]. Among the two, the low-energy high conductance stripe involves the transition from $|D_\uparrow\rangle$ to $|T_+^*\rangle$ and thus tunneling of spin-↑ electrons through the QD, while the high-energy high conductance stripe involves the transitions of $|D_\uparrow\rangle \rightarrow |S^*\rangle$ and $|D_\uparrow\rangle \rightarrow |T_0^*\rangle$ and tunneling of spin-↓ electrons through the QD. Since the three two-electron excited states consist of an electron of the same spin (the ↑ spin in this case) in the first quantum level and an electron of different spins in the second quantum level in the QD, the energy difference between the two high conductance stripes at a given magnetic field $B_X$ is a measure of the Zeeman splitting $\Delta E(B_X) = |g_{IIX}^* \mu_B B_X|$ of the second quantum level (level $II$) in the QD, where $g_{IIX}^*$ is the effective electron g-factor of a spin-1/2 electron at the second quantum level. Figure 2(d) displays the energy differences (data points), extracted from the peak values of the two high conductance stripes shown in figure 2(c), as a function of $B_X$. Through a line fit (red solid line) to the data in figure 2(d), an effective electron g-factor of $g^*_{IIX} \sim 50.2$ is extracted. Similar measurements are performed with the magnetic field applied along the Y and Z directions and the corresponding g-factors of $g^*_{IIY} \sim 46.7$ and $g^*_{IIZ} \sim 47.9$ are extracted for the second quantum level in the QD, see the inset of figure 2(d) for a summary of the three g-factor values extracted for the second quantum level. We find that the g-factor of the second quantum level is again anisotropic and is different from the g-factor of the first quantum level in the QD.

*3.3. g-factor tensors of the first two quantum levels of the QD in device A*

The above measurements demonstrate that the extracted g-factor of a spin-1/2 electron in the InSb nanowire QD is sensitively dependent on the energy level and on the direction of the applied magnetic field. Thus, in general, the g-factor of a spin-1/2 electron in a quantum level *m* of the InSb nanowire QD should be described by a tensor $g^*_m$ and its value determined under an applied magnetic field **B** of an arbitrary direction can be expressed as [8,9]

$$|g^*_m(\mathbf{B})| = \frac{\sqrt{g_1^2 B_1^2 + g_2^2 B_2^2 + g_3^2 B_3^2}}{|\mathbf{B}|}, \qquad (1)$$

where $g_1$, $g_2$ and $g_3$ are the three principal values of the g-factor tensor defined with respect to the three orthogonal principal axis directions, and $B_1$, $B_2$ and $B_3$ are the magnetic field components along these directions. We emphasize that these principal axis directions are generally different from the axis directions in the coordinate system we have set above in the measurements. However, the principal axis coordinate system of the g-factor tensor can be transformed from our measurement coordinate system by means of the three successive rotations defined through the Euler angles of rotation (α, β, γ). To determine the principal values and the orientations of the principal axes of the g-factor tensor $g^*_m$ in our InSb nanowire QD, we perform the measurements for the values of $g^*_m$ with the rotations of the magnetic field **B** in three orthogonal planes and fit the measured values to Eq. (1). These three orthogonal planes are defined as the X-Y, Z-X, and Z-Y planes as described by the three schematics shown in figure 3.

Figures 3(a)-3(c) show the results of such measurements for the g-factor tensor

$g_I^*$ of a spin-1/2 electron in the first quantum level of the InSb nanowire QD in device A. Here, the two current peaks, corresponding to the resonant transport through the Zeeman splitting states of the first quantum level, are found in the measurements of current $I_{ds}$ as a function of $V_{bg}$ at $V_{ds} = 0.2$ mV under rotations of applied magnetic field **B** in the three orthogonal X-Y, Z-X, and Z-Y planes. The strength of the magnetic field is set at |**B**| = 0.4 T, which is within a range of magnetic fields in which the Zeeman splitting is expected to show a good linear dependence on |**B**| in all the magnetic field directions (see experimental observations as shown in figure 2(b) for example). The extracted energy difference between the two Zeeman splitting states from the measurements is given by $\Delta E(\mathbf{B})=E_C+g_I^*\mu_B|\mathbf{B}|$. Figure 3(a) displays the experimentally extracted values of $g_I^*$ with **B** applied in the X-Y plane. It is found that $g_I^*$ oscillates with the rotation of **B** with its maxima appearing approximately at rotation angles of $\phi \sim 0°$ and 180° and its minima appearing approximately at $\phi \sim 90°$ and 270°. Oscillations of $g_I^*$ with the rotation of **B** are also found in figures 3(b) and 3(c), where **B** is applied in the Z-X and Z-Y planes. However, in difference from the case shown in figure 3(a), the maximum (minimum) values of $g_I^*$ do not appear at about 0° and 180° (90° and 270°) in figures 3(b) and 3(c). The red solid curves in figures 3(a)-3(c) are the fit of the measured $g_I^*$ values to Eq. (1) with the principal values ($g_1$, $g_2$, $g_3$) and the Euler angles (α, β, γ) as free fitting parameters. The fit yields ($g_1$, $g_2$, $g_3$) = (60.5, 50.1, 56.5) and (α, β, γ) = (96.4°, 71.7°, 155.6°). Clearly, it is seen that the effective g-factor tensor in the QD is spatially anisotropic and is level dependent. In the absence of SOI, the g-factor in a symmetric QD should be isotropic. The anisotropy of the $g_I^*$ tensor observed here can thus be attributed to the complex profile of the quantum confinement in the InSb nanowire QD and to anisotropic orbital contributions [17,25] due to the presence of strong SOI in the InSb quantum structure [7] as we will discuss below. For the principal axis directions of the $g_I^*$ tensor, an intuitive view is to present them in terms of the polar angle $\vartheta_i$ with respect to the Z axis and the azimuth angle $\varphi_i$ defined in the X-Y plane with respect to the X direction of the measurement coordinate system. Here, subscript $i$=1, 2 or 3 is the index for a principal axis. For the $g_I^*$ tensor of the first quantum level in the QD of device A, the orientations of the three principal axes can be mapped out from the obtained Euler angles of rotation of (α, β, γ) = (96.4°, 71.7°, 155.6°) as ($\vartheta_1$, $\varphi_1$)=(66.9°, 1.7°), ($\vartheta_2$, $\varphi_2$)=(108.3°, 83.6°) and ($\vartheta_3$, $\varphi_3$)=(30.1°, 138.9°), respectively. These results show that the first principal axis of the $g_I^*$ tensor with the largest principal value of $g_1 \sim 60.5$ lies approximately in the Z-X plane and points to a direction more parallel to the nanowire axis. The other two principal axes with two smaller principal values point to directions which are nearly perpendicular to the nanowire axis. These results are consistent with the fact that the cross-sectional confinement in the nanowire is relatively weaker than that along the nanowire axis, due to a strong depletion effect of the Schottky barriers in the last few-electron regime in the nanowire QD [10,14,37].

Figures 4(a)-4(c) show the extracted values of $g_{II}^*$ i.e., the g-factor values for the second quantum level of the QD with the magnetic field applied in the X-Y, Z-X, and Z-Y planes, respectively. Here, the Zeeman splitting energies of the quantum level of the QD are extracted from the excitation spectra measured along the line cut A in figure 1(c) at a fixed magnetic field magnitude of |**B**| = 0.8 T, which is within a range of 0 to 0.9 T in which the magnetic field evolution of the Zeeman energy is, to a good approximation, linear in |**B**| for all magnetic field directions [see, for example, a result shown in figures 2(c) and 2(d)]. The red solid curves in figures 4(a)-4(c) are the results of the fit of the measurements to Eq. (1). The $g_{II}^*$ tensor extracted from the fit has the three principal values $(g_1, g_2, g_3) = (50.7, 46.0, 47.8)$ and the Euler angles of rotation $(\alpha, \beta, \gamma) = (96.1°, 97.4°, 2.1°)$ of the principal axes with respect to the measurement coordinate system. The directions of the three principal axes viewed with respect to the measurement coordinate system are found to be $(\vartheta_1, \varphi_1)=(87.9°, 8.1°)$, $(\vartheta_2, \varphi_2)=(96.1°, 97.8°)$ and $(\vartheta_3, \varphi_3)=(6.5°, 117.2°)$, respectively. Again, here we see that $g_{II}^*$ is anisotropic and its first principal axis with a principal value of $g_1 \sim 50.7$ points to a direction close to be parallel to the nanowire axis. However, when comparing the results obtained for $g_I^*$ and $g_{II}^*$ tensors, we find that both the principal values and the principal axes are level dependent, similar to the results reported previously for the g-factors in other quantum structures [9,14]. For convenience, we list the obtained orientations of the principal axes of the $g_I^*$ and $g_{II}^*$ tensors together with their corresponding principal values in Table I.

3.4. *Measurements of g-factors and SOI for the QD in device B*
We have also studied device B made from a nanowire with a slightly smaller diameter and obtained similar results for the g-factor as in device A. Figure 5(a) shows the differential conductance $dI_{ds}/dV_{ds}$ of the InSb nanowire QD in device B measured as a function of source-drain bias voltage $V_{ds}$ and back-gate voltage $V_{bg}$. Here, several well-defined Coulomb blockade diamond structures are observed. Since no more Coulomb diamond structure is observable at back-gate voltages $V_{bg}$ lower than −1.35 V, the QD is empty of electrons at $V_{bg}$ below −1.35 V. The electron number in each Coulomb blockade region can be assigned and is again marked by an integer *n*. At *n* = 2, 4, and 6, the first, second and third quantum levels (labelled as quantum levels *I*, *II*, and *III*) are successively fully filled with electrons in the QD. Figure 5(b) shows the current $I_{ds}$ of the device measured at $V_{ds}$ = 0.2 mV as a function of $V_{bg}$ and magnetic field $B_Z$ applied along the Z direction. Three pairs of high current stripes are seen in the figure and each pair corresponds to electron occupations of the two spin states of a quantum level in the QD. Here each spin state is again marked by an arrow. Figure 5(c) shows the extracted spin splitting energies of the three quantum levels as a function of $B_Z$. By line fits to the measurement data, the values of the g-factor $g^*_{IZ}$~52.5, $g^*_{IIZ}$~40.3 and $g^*_{IIIZ}$~53.0 are obtained for quantum levels *I*, *II* and *III*, respectively. Similar measurements are performed with magnetic fields $B_X$ and $B_Y$ applied along the X and Y axes and the corresponding g-factor values of $g^*_{IX}$~57.4, $g^*_{IIX}$~47.5 and $g^*_{IIIX}$~62.4, and $g^*_{IY}$~50.3, $g^*_{IIY}$~43.1 and $g^*_{IIIY}$~53.7 are extracted. These results show that the g-factor of a spin-1/2 electron in the InSb nanowire QD of device B also is sensitively dependent on quantum level and is spatially anisotropic.

Finally, we briefly demonstrate, using device B, the presence of strong SOI in our InSb nanowire QDs. Figure 5(d) displays the differential conductance $dI_{ds}/dV_{ds}$ measured along the line cut B in figure 5(a) as a function of $V_{ds}$ and $B_Z$. The three high conductance stripes are found in the figure. On the low field side, these three high conductance stripes, from top to bottom, involve the transitions from the spin-↑ doublet ground state $|D_\uparrow\rangle$ to the two-electron singlet ground state $|S\rangle$, to the two-electron triplet excited state $|T^*_+\rangle$, and to the two-electron singlet and triplet excited states $|S^*\rangle$ and $|T^*_0\rangle$ in the QD, as marked in figure 5(d). With increasing $B_Z$, the top high conductance stripe involving the transition from state $|D_\uparrow\rangle$ to state $|S\rangle$ shifts towards higher $|V_{ds}|$. The other two high conductance stripes move apart and the one involving the transition from state $|D_\uparrow\rangle$ to state $|T^*_+\rangle$ is found to shift towards lower $|V_{ds}|$ with increasing $B_Z$, leading to an avoided crossing with the top high conductance stripe at $B_Z$~0.9 T due to the presence of SOI. In order to extract the strength of SOI, $\Delta_{so}$, seen at this experiment, we fit the energy positions of the quantum states extracted from the two anti-crossed high conductance stripes to a simple two-level perturbation model [36]

$$E_\pm = \frac{E_S + E_{T_+^*}}{2} \pm \sqrt{\frac{(E_S - E_{T_+^*})^2}{4} + \Delta_{so}^2}, \tag{2}$$

where $E_S$ and $E_{T_+^*}$ are the energy levels of the singlet state $|S\rangle$ and the triplet state $|T_+^*\rangle$ without taking SOI into account. Figure 5(e) displays the energy positions (data points) extracted from the two anti-crossed high conductance stripes shown in figure 5(d), while the solid lines in figure 5(e) are the fit. The fit yields a large value of $\Delta_{so} \sim 180$ μeV, consistent with the results reported previously [7]. The values of the g-factors, $g_{IZ}^* \sim 53.6$ and $g_{IIZ}^* \sim 42.9$, are also obtained from the fit. These values are in good agreement with the values extracted above from the measurements shown in figures 5(b) and 5(c).

## 4. Conclusions

In summary, we have measured the Zeeman splitting energies of electrons in single InSb nanowire QDs under different, three-dimensionally oriented magnetic fields. The measurements show that the extracted g-factor in the QDs remains large but varies from level to level. The measurements also show that the g-factor in the QDs is spatially anisotropic and should be described by a tensor for completeness. We have demonstrated the experimental determination of the g-factor tensors in an InSb nanowire QD by the Zeeman energy measurements of a spin-1/2 electron at quantum levels of the QD under rotations of a magnetic field in three orthogonal planes. The anisotropy and the level dependence of the g-factor can be attributed to the presence of strong SOI and asymmetric confining potentials in the QDs. In this work, a SOI strength of $\Delta_{so} \sim 180$ μeV is observed in the measurements of the magnetic field evolution of the excitation spectra of an InSb nanowire QD. Our work provides a first solid demonstration for the complex nature of the g-factor in InSb nanowire quantum structures. Such information and its determination play a crucial role in successful realization and accurate manipulation of spin qubits and topological quantum states made from such semiconductor nanowire structures.

**Acknowledgements**
We thank Philippe Caroff for the growth of the InSb nanowires employed in this work.

This work is supported by the Ministry of Science and Technology of China through the National Key Research and Development Program of China (Grant Nos. 2017YFA0303304, 2016YFA0300601, 2017YFA0204901, and 2016YFA0300802), the National Natural Science Foundation of China (Grant Nos. 91221202, 91421303, 11874071 and 11974030), the Beijing Academy of Quantum Information Sciences (No. Y18G22), and the Beijing Natural Science Foundation (Grant No. 1202010).

**CAPTIONS**

**Figure 1.** (a) SEM image (in false color) of a representative device made on a Si/SiO$_2$ substrate. A single QD is defined in an InSb nanowire between the two Ti/Au metal contacts via naturally formed Schottky barriers. (b) Cross-sectional schematic view of the device displaying the heavily n-doped Si substrate (grey) capped with a 200-nm-thick layer of SiO$_2$ (maroon) on top, the nanowire (green), and the two metal contacts (orange), and measurements circuit setup. The measurement coordinate system is shown in the lower-left corner, in which the X axis points along the nanowire axis, the Z axis is set to be perpendicular to the substrate plane, and the Y axis is set to be in-plane, perpendicular to the nanowire axis. (c) Differential conductance $dI_{ds}/dV_{ds}$ measured for device A (see the text for the description of the device) as a function of source-drain bias voltage $V_{ds}$ and back-gate voltage $V_{bg}$ (charge stability diagram). Integer numbers $n$ mark the numbers of electrons in the QD formed in the device.

**Figure 2.** (a) Current $I_{ds}$ measured for device A at $V_{ds} = 0.2$ mV as a function of back-gate voltage $V_{bg}$ and magnetic field $B_X$ applied along the X direction. The two high conductance stripes labeled with ↑ and ↓ result from electron tunneling through the spin-↑ and spin-↓ states of the first quantum level in the QD. (b) Energy differences between the two spin states (data points) extracted from the measurements in (a). By a line fit to the data, a value of the g-factor of $g^*_{IX} \sim 58.5$ for the first quantum level is obtained. The inset shows the orientation of the nanowire and the three extracted values of the g-factor at magnetic fields applied along the X, Y and Z axes for the first quantum level. (c) Differential conductance $dI_{ds}/dV_{ds}$ measured along the line cut A in figure 1(c) as a function of $V_{ds}$ and magnetic field $B_X$ applied along the X direction. The three high conductance stripes, from top to bottom, involve the quantum state transitions from the spin-↑ doublet ground state $|D_\uparrow\rangle$ to the two-electron singlet ground state $|S\rangle$, to the two-electron excited state $|T^*_+\rangle$, and to the two-electron excited states $|S^*\rangle$ and $|T^*_0\rangle$, respectively. (d) Energy differences extracted from the peak values of the two high conductance stripes in (c). By a line fit to the data, a value of the g-factor $g^*_{IIX} \sim 50.2$ for the second quantum level is obtained. The inset shows the extracted values of the g-factor at magnetic fields applied along

the X, Y and Z axes for the second quantum level.

**Figure 3.** (a) Values of $g_I^*$ extracted from the Zeeman splitting of the first quantum level of the QD in device A, measured in a similar way as in figures 2(a) and 2(b), as a function of the rotation angle of a magnetic field **B** with |**B**|=0.4 T in the X-Y plane as indicated in the schematics (top panel). (b) and (c) The same as in (a) but for the magnetic field rotated in the Z-X and Z-Y planes. The red solid lines are the results of the fit of the experimental data to Eq. (1).

**Figure 4.** (a)-(c) Values of $g_{II}^*$ extracted from the Zeeman splitting of the second quantum level of the QD in device A, measured in a similar way as in figures 2(c) and 2(d), (i.e., from the excitation spectra), as a function of the rotation angle of magnetic field **B** with |**B**|=0.8 T in the X-Y, Z-X, and Z-Y planes as indicated in the schematics in figure 3. The red lines are the results of the fit of the experimental data to Eq. (1).

**Figure 5.** (a) Differential conductance $dI_{ds}/dV_{ds}$ measured for device B (see the text for the description of the device) as a function of back-gate voltage $V_{bg}$ and source-drain bias voltage $V_{ds}$ (charge stability diagram). Integer numbers $n$ mark the numbers of electrons in the QD. (b) Current $I_{ds}$ measured for device B at $V_{ds} = 0.2$ mV as a function of back-gate voltage $V_{bg}$ and magnetic field $B_Z$ applied along the Z direction. Three pairs of high conductance stripes correspond to electron transport through the spin states of the first three quantum levels (labelled as *I*, *II* and *III*) in the QD. Corresponding spin fillings of the quantum levels are indicated with arrows in the plot. (c) Energy differences of the two spin states extracted from the measurements for quantum levels *I*, *II* and *III* as shown in (b) as a function of $B_Z$. By line fits to the data points, the g-factor values of $g_{IZ}^* \sim 52.5$, $g_{IIZ}^* \sim 40.3$ and $g_{IIIZ}^* \sim 53.0$ are obtained for quantum level *I*, *II* and *III*, respectively. (d) Differential conductance $dI_{ds}/dV_{ds}$ measured along line cut B in (a) as a function of $V_{ds}$ and $B_Z$. The three high conductance stripes, from top to bottom, involve the quantum state transitions from the spin-↑ doublet ground state $|D_\uparrow\rangle$ to the two-electron singlet ground state $|S\rangle$, to the two-electron excited state $|T_+^*\rangle$, and to the two-electron excited states $|S^*\rangle$ and $|T_0^*\rangle$. (e) Converted energy positions at the peak values of the two anti-crossed high conductance stripes in (d). The red solid lines show the fit of the data points to a two-level perturbation model and $\Delta_{so}$ stands for the strength of SOI seen in the experiment.

**Table I**. Principal values ($g_1$, $g_2$, $g_3$) and orientations [($\vartheta_1$, $\varphi_1$), ($\vartheta_2$, $\varphi_2$), ($\vartheta_3$, $\varphi_3$)] of the principal axes of g-factor tensors $g_I^*$ and $g_{II}^*$ determined for the first and second quantum levels in the InSb nanowire QD of device A.

| Quantum Level | $g_1$ | $g_2$ | $g_3$ | $\vartheta_1$ | $\varphi_1$ | $\vartheta_2$ | $\varphi_2$ | $\vartheta_3$ | $\varphi_3$ |
|---|---|---|---|---|---|---|---|---|---|
| I | 60.5 | 50.1 | 56.5 | 66.9° | 1.7° | 108.3° | 83.6° | 30.1° | 138.9° |
| II | 50.7 | 46.0 | 47.8 | 87.9° | 8.1° | 96.1° | 97.8° | 6.5° | 117.2° |

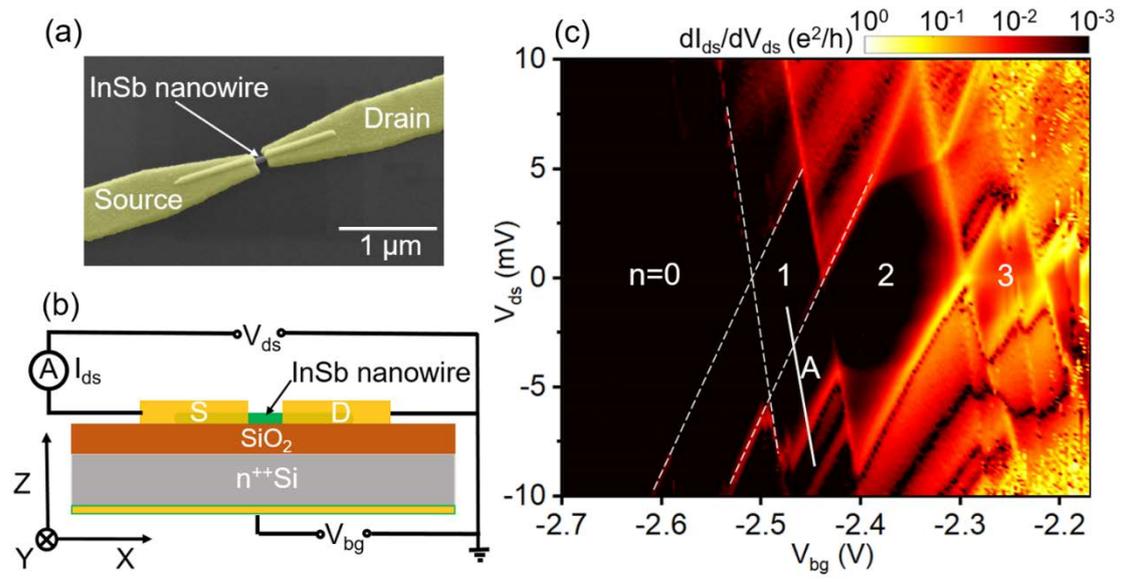

**Figure 1, Jingwei Mu *et al*.**

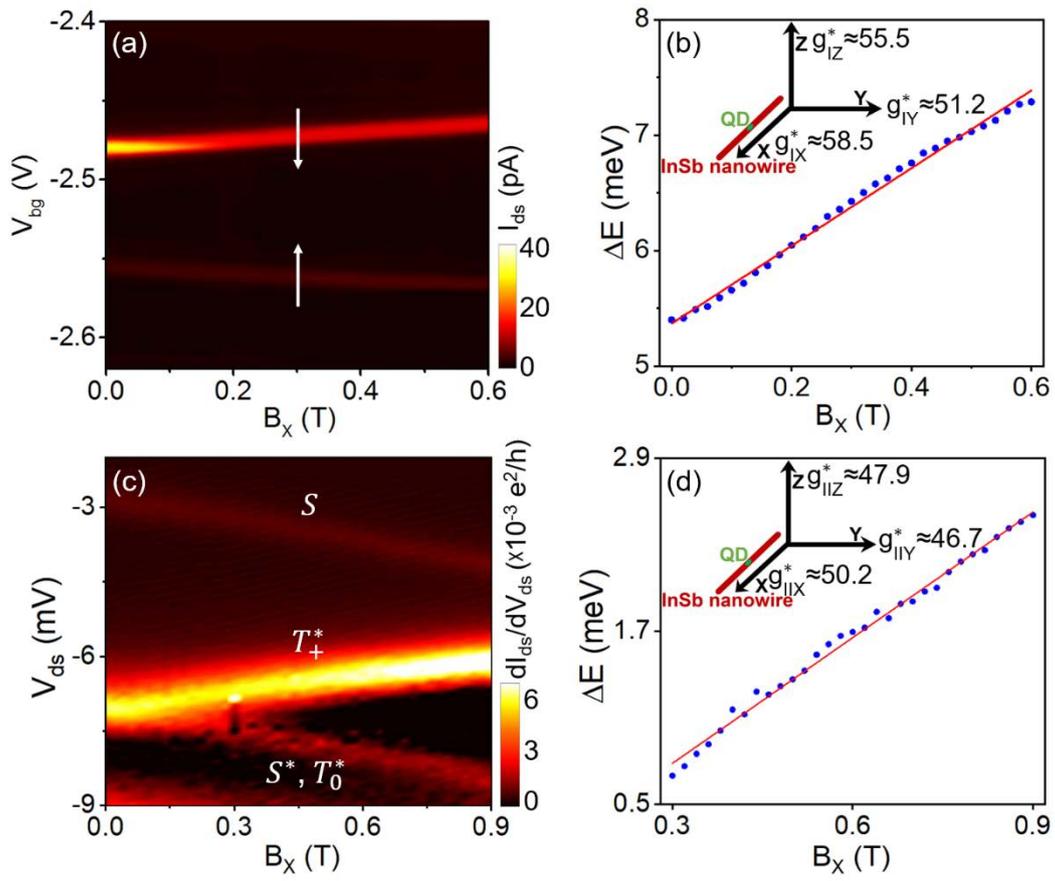

**Figure 2, Jingwei Mu *et al*.**

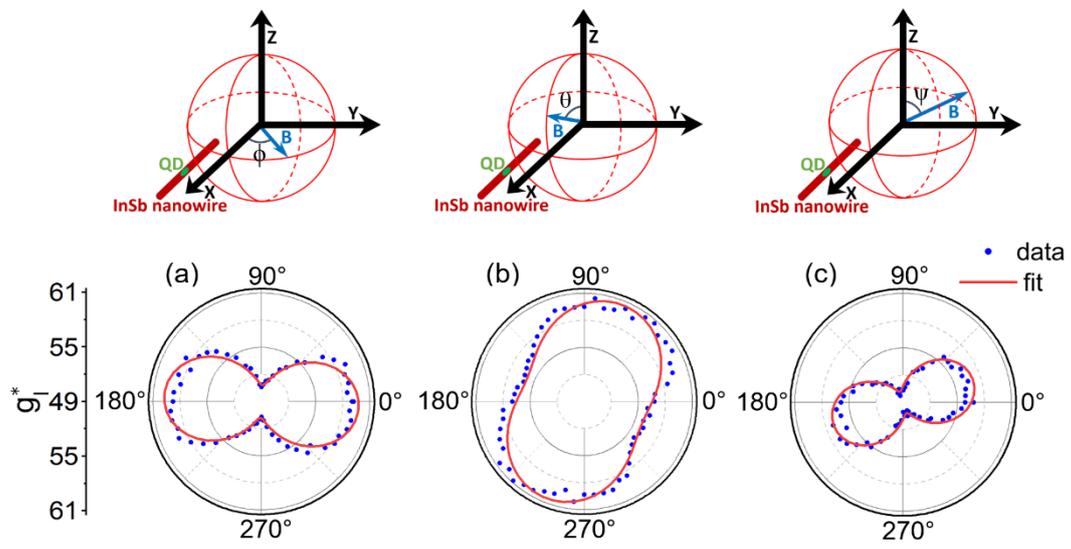

**Figure 3, Jingwei Mu *et al*.**

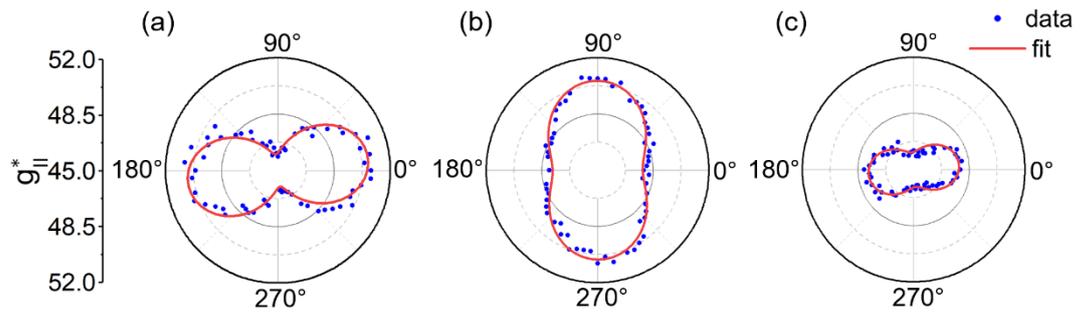

**Figure 4, Jingwei Mu** *et al*.

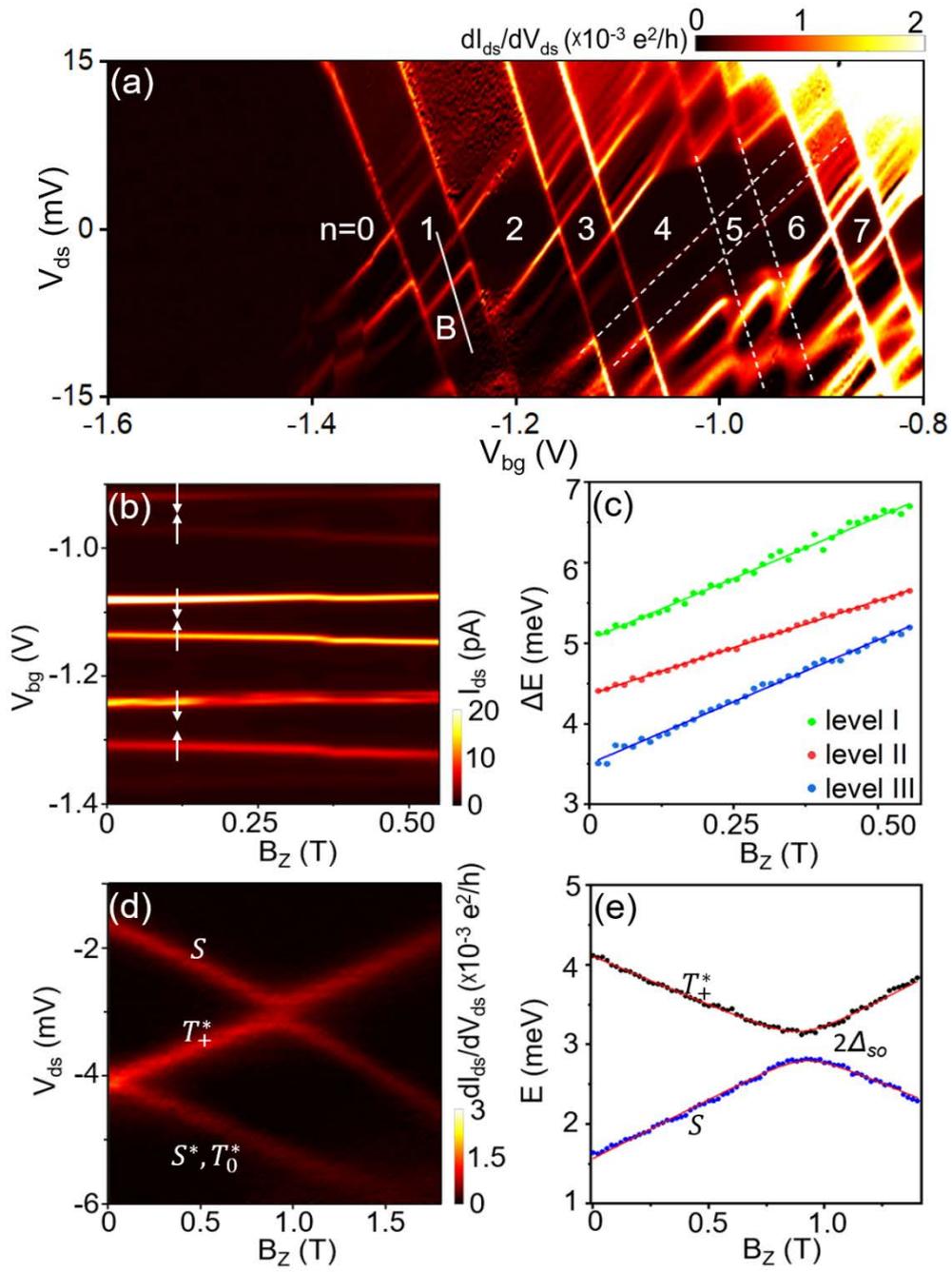

**Figure 5, Jingwei Mu *et al*.**